\documentclass[aps,prd,floatfix]{revtex4}
\usepackage{graphicx}
\usepackage{dcolumn}
\usepackage{bm}
\usepackage{amssymb,hyperref,amsmath}

\newcommand{\be}{\begin{equation}}
\newcommand{\ee}{\end{equation}}
\newcommand{\A}{{\rm A}}
\newcommand{\M}{{\rm M}}
\newcommand{\tpz}{\ensuremath{^3\textrm{P}_0\;}}
\newcommand{\tso}{\ensuremath{^3\textrm{S}_1\;}}
\newcommand{\tdo}{\ensuremath{^3\textrm{D}_1\;}}

\newcommand{\ps}    {\ensuremath{ \psi(4415) }}
\newcommand{\dt}    {\ensuremath{ \overline D{}^{*}_2(2460)   }}
\newcommand{\ddt}   {\ensuremath{ D  \overline D{}^{*}_2(2460)   }}
\newcommand{\dpi}   {\ensuremath{ D^-   \pi^+ }}

\begin{document}

\title{Update on Charmonium Theory}

\author{T.Barnes\footnote{Email: tbarnes@utk.edu}}

\affiliation{
Physics Division, Oak Ridge National Laboratory,
Oak Ridge, TN 37831-6373, USA\\
Department of Physics and Astronomy, University of Tennessee,
Knoxville, TN 37996-1200, USA}

\begin{abstract}
In this invited presentation I review some recent developments in the 
theory of charmonium that appear likely to be of importance for future 
experimental studies in this field. The specific areas considered are 
double charmonium production, LQCD studies of charmonium, recent
results for hadron loops, $c\bar c$ production cross sections at PANDA,
charm molecules, and two recent developments, ``charmiscelleny". 
\end{abstract}

\maketitle


\begin{center}
{\bf Introduction}
\end{center}

Charmed hadron spectroscopy has undergone a renaissance in recent years, 
due to an impressive increase in experimental results, 
which are often in striking disagreement with theoretical expectations regarding these hadrons.
This renaissance began with the discovery 
of the surprisingly light charm-strange mesons~\cite{Aubert:2003fg,Aubert:2003pe},
and has continued with the observation of the charm-meson molecule candidate 
X(3872) and a ``zoo" of latter-alphabet X,Y and Z states,
which include conventional 2P $c\bar c$ candidates, hybrid charmonium candidates, and 
``charged charmonia", which are of course charmed meson molecule candidates. 
For recent reviews of the experimental status of this field, see 
Refs.\cite{Swanson:2006st,Olsen:2009zz}. Many of these 
exciting experimental discoveries have rather surprisingly originated from facilities 
that were not originally intended for research in this area, notably 
the $e^+e^-$ ``B factories", which have proven to be copious sources of charmed hadrons.

In this report I will discuss recent developments in our understanding of the theory that 
underlies charmonium states and related topics. My criteria in selecting material for this talk 
are that the results should be clearly relevant to experiment, reasonably recent, and likely to 
be relevant to future experimental studies in this field. The specific topics chosen for 
review are as follows:
\begin{enumerate}
\item
Double charmonium production 
\item
LQCD studies of charmonium 
\item
Recent results for hadron loops 
\item
Charmonium production cross sections at PANDA
\item
Charm molecules 
\item
``Charmiscelleny"; two recent developments. 
\end{enumerate}
\vskip 1cm

\section{Double Charmonium Production}

\begin{figure}[h]
  \includegraphics[width=2.5in]{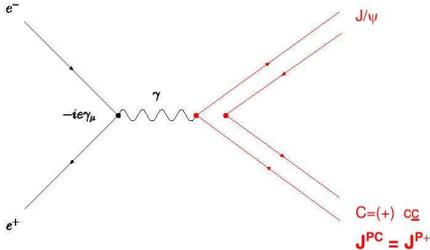}
  \caption{\label{fig:fig1} Double charmonium production in $e^+e^-$ annihilation.}
\end{figure}

One of the interesting new experimental techniques that has been exploited recently in
charmonium studies is ``double charmonium production", in which $e^+e^-$ annihilation
produces two charmonium resonances (see Fig.\ref{fig:fig1}). In the results reported thus far,
a $J/\psi$ has been used as a trigger; the missing mass of the recoiling system shows clear peaks 
at the masses of the $\eta_c$, $\chi_0$ and $\eta_c'$, and a higher-mass enhancement referred 
to as the X(3943) is also observed.   
(See for example results reported by the Belle 
\cite{Abe:2004ww,Abe:2007jn}
and
BABAR \cite{Aubert:2005tj}
collaborations.)

The double charmonium production process is especially interesting in that it gives us access 
to $J^{PC} \ne 1^{--}$ $c\bar c$ states at $e^+e^-$ machines. 
This includes C~=~(+) $c\bar c$ mesons in particular, which are otherwise rather difficult 
to produce; typically they have been studied using hadronic production, as in $p\bar p$ 
annihilation at Fermilab, in 
radiative transitions from higher-lying $J^{PC} = 1^{--}$ states, or in the two-photon process
$e^+e^- \to e^+e^- \gamma\gamma, \gamma\gamma \to c\bar c$, 
which has rather small cross sections.  

\eject

\begin{figure}[h]
  \includegraphics[width=3in]{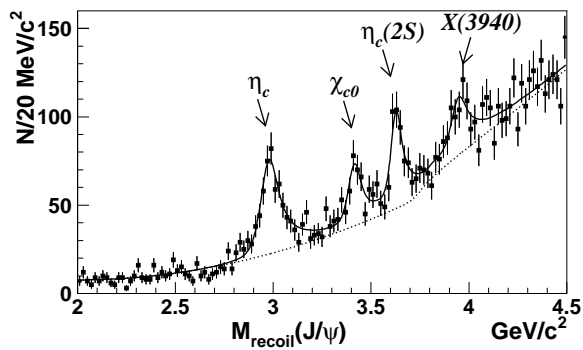}
  \caption{\label{fig:Belle_fig1_new} States seen in double charmonium production 
(recoiling against the $J/\psi$)
by Belle \cite{Abe:2007jn}.}
\end{figure}

Although Fig.\ref{fig:Belle_fig1_new} 
shows that this is clearly a successful approach for producing C~=~(+) charmonia in $e^+e^-$
annihilation, it would be even more interesting if the double charmonium production mechanism 
were well understood theoretically; in this case one could ideally predict the production
cross sections for various $c\bar c$ states, and non-$c\bar c$ candidates could be identified 
through their anomalous cross sections. In addition one might hope to use this mechanism to
observe the as yet unreported $J^P=2^-$ $c\bar c$ D-wave states $\eta_{c2}$ and $h_{c2}$; these are 
expected to be quite narrow, due to the absence of open-charm hadronic decay modes.   

The theoretical mechanism of double charmonium production has been studied in some detail 
by three groups, at ANL, Beijing and Oxford. In their important initial study the ANL group
(Braaten and Lee \cite{Braaten:2002fi}) 
assumed a nonrelativistic one-gluon-exchange (OGE) mechanism for production of the 
``second" $c\bar c$ pair. 
(One might {\it a priori} have anticipated important 
contributions either from pQCD-type mechanisms such as OGE, or from the nonperturbative 
decay mechanisms that appear to dominate light hadron strong decays, as are described by the 
\tpz model.) Braaten and Lee concluded that the OGE decay mechanism gave a characteristic 
large amplitude for 
$e^+e^- \to J/\psi + \chi_{c0}$ relative to the other $\chi_{cJ}$ states, with approximate 
relative cross sections of
\be
\sigma(e^+e^- \to J/\psi + (\chi_{c2} : \chi_{c1} : \chi_{c0})) \approx 
\ 3\; :\; 2\; :\; 10\; .
\ee 
This explains the dominance of the $\chi_{c0}$ over the other 1P states that is 
observed in the data (see for example Fig.\ref{fig:Belle_fig1_new}). 
However it was also noted that these leading order (OGE) NR pQCD 
cross sections are about an order of magnitude smaller than the experimental results.

As this was a rather complicated pQCD calculation, it was clearly important to have an
independent check of these results. This work was confirmed and extended 
in a series of papers by the Beijing group of 
Chao, Liu, He, Ma and Zhang; see Ref.\cite{Liu:2004ga} 
and references cited therein. Liu, He and Chao \cite{Liu:2004ga} have recently  
reported very interesting predictions of various double charmonium cross sections, including
results for states that have not yet been observed, such as the narrow $\eta_{c2}$ and $h_{c2}$ states.
Their results for these cross sections (relative to the reference channel $J/\psi + \eta_c$) 
are shown in Table~\ref{table:LHC_dccbar_ratios}; 
evidently these predictions imply that the double charmonium production
mechanism has considerable potential for the discovery of new states, given an improvement 
over current statistics by one to two orders of magnitude.

\begin{table}[h]
\begin{tabular}{|l|l|l|l|l|l|l|l|l|c|}
\hline
&$\eta_{c}(1S,2S,3S)$&$\chi_{c0}(1P,2P)$&$\chi_{c1}(1P,2P)$&$\chi_{c2}(1P,2P)$&$h_{c}(1P,2P)$
&$^3D_{1}$& $^3D_{2}$&$^3D_{3}$&$\eta_{c2}({}^1D_2)$
\\
\hline
$\psi$(1S)&1.0,0.65,0.56&1.05,1.4&0.15,0.21&0.27,0.36&&&&&0.03
\\
\hline
$\psi(2S)$&0.65,0.42,0.36&0.68,0.94&0.11,0.14&0.17,0.24&&&&&0.02
\\
\hline $\eta_{c}$(1S)&&&&&0.11,0.15&0.025&&&
\\
\hline $\eta_{c}$(2S)&&&&&0.07,0.10&0.016&&&
\\
\hline $\eta_{c}$(3S)&&&&&0.06,0.08&&&&
\\
\hline $\chi_0$(1P)&&&&&0.03,0.05&&&&
\\
\hline $\chi_1$(1P)&&&&&0.15,0.21&0.012&0.015&0.006&
\\
\hline $\psi_2$(1P)&&&&&0.010,0.013&0.006&0.001&0.0015&
\\
\hline
\hline
\end{tabular}
\caption{\label{table:LHC_dccbar_ratios}Double charmonium production cross sections 
relative to $J/\psi+\eta_c$ at $\sqrt{s}=10.6$~GeV
found by Liu {\it et al.}~\cite{Liu:2004ga}}
\end{table}

Regarding the problematic overall scale of these cross sections, Zhang, Ma and Chao 
\cite{Zhang:2008gp} considered NLO corrections to the OGE model, and found that these 
considerably reduced the reported discrepancy between the observed and calculated 
double charmonium production cross sections. Thus the well-known problem of the scale of
these cross sections may have been due to the neglect of higher-order pQCD effects.

The Oxford group of Close and Downum \cite{Close:2008xn} has also investigated double 
charmonium production. They too confirmed the NR pQCD OGE calculation of Braaten and Lee 
analytically, and have in addition considered nonperturbative decay mechanisms 
of \tpz type (see Burns, Close and Thomas \cite{Burns:2007hk}). Close and Downum
found that in the latter case the final state $J/\psi \chi_{c0}$ would no longer be the 
dominant $J/\psi \chi_{cJ}$ mode, which argues in favor of pQCD processes for double 
charmonium production. They note that in contrast to the double charmonium results, 
in the light meson sector one sees vector + tensor dominance, as in $e^+e^- \to \omega f_2$,
which is a prediction of the nonperturbative decay mechanism. 

\section{LQCD Charmonium}

Here I will be very succinct, as the very important new work in this area will be discussed 
in detail in a dedicated plenary talk by C.Thomas~\cite{Thomas}.  

Although LQCD studies of charmonium spectroscopy have existed in the literature for some time
(see for example Ref.\cite{Liao:2002rj}), these studies have not had especially high impact 
because they have generally confirmed the predictions of $c\bar c$ potential models and 
experimental results for the known spectrum of $c\bar c$ 
states. More dramatic early predictions from LQCD have included mass estimates for quarkonium hybrids,
such as {\it ca.}~4.4~GeV for the lightest $1^{-+}$ exotic charmonium hybrid. 

Recently however very impressive new results on charmonium from LQCD have been reported,
notably by the JLAB lattice QCD collaboration of Dudek, Edwards, Peardon, Richards and Thomas 
\cite{Dudek:2009kk,Dudek:2009qf}. The most interesting 
results for charmonium involve determinations of the photocouplings of charmonium sector resonances,
{\it including} $J^{PC}$-exotic charmonium hybrids \cite{Dudek:2009kk}.
The importance of these new results for the JLAB experimental program would be hard to overstate. 
The 12~GeV upgrade at JLAB and the Hall-D GlueX experiment 
in particular are predicated on the hope that the photocouplings of hybrid mesons are sufficiently 
large for the spectrum of these states to be clearly established through a high-statistics study
of meson photoproduction. Although models of hybrids suggest that their photocouplings 
should not be small, these of course make assumptions regarding the nature of hybrids that 
are conjectural, and may be inaccurate. Without strong theoretical evidence of large hybrid 
photocouplings, especially for the ``smoking-gun" $J^{PC}$-exotics, the plan to establish these 
states experimentally using photoproduction at JLAB is not well justified theoretically. 
Clearly, lattice QCD studies 
of the photocouplings of hybrids have an extremely important role to play in establishing 
the sensitivities required for a definitive study of hybrids using photoproduction reactions,
due to their better controlled systematic uncertainties.

\begin{table}[ht]
 \begin{tabular}{c c| c   c | cc}
$\begin{matrix}
\mathrm{sink} \\ 
\mathrm{level}
\end{matrix}$ 
& $\begin{matrix}
\mathrm{suggested} \\ 
\mathrm{transition}
\end{matrix}$  
& $\hat{V}(0)$ 
& $\begin{matrix}
\beta/\mathrm{MeV} \\ 
\lambda/\mathrm{GeV^{-2}} 
\end{matrix}$
& $\Gamma_{\mathrm{lat}}$/keV 
& $\Gamma_{\mathrm{expt}}$/keV \\
\hline
0 & $J/\psi \to \eta_c \gamma$ 
& $1.89(3)$ 
& $\begin{matrix} 
513(7) \\ 
\mathrm{0[fixed]}
\end{matrix}$ 
& $2.51(8)$ 
& $1.85(29)$ \\
1 & $\psi' \to \eta_c \gamma$ 
& $0.062(64)$ 
& $\begin{matrix} 
530(110) \\ 
4(6)
\end{matrix}$ 
& $0.4(8)$ 
& $\begin{matrix}
0.95(16)\\ 
1.37(20)
\end{matrix}$ \\
3 & $\psi'' \to \eta_c \gamma$ 
& $0.27(15)$ 
& $\begin{matrix} 
367(55) \\ 
-1.25(30)
\end{matrix}$ 
& $10(11)$ 
& - \\
5 & $Y_{\mathrm{hyb.}} \to \eta_c \gamma$ 
& $0.28(6)$ 
& $\begin{matrix} 
250(200) \\ 
\mathrm{0[fixed]}
\end{matrix}$ 
& $42(18)$ 
& -
\end{tabular}
\caption{\label{table:Dudek_LQCD}M1 radiative decay widths for charmonium and related states from LQCD, as reported
by Dudek {\it et al.}~\cite{Dudek:2009kk}.}
\end{table}

An example of the very important new results reported by the JLAB LQCD group, 
from the viewpoint of using photoproduction to identify hybrids, appears 
in their table of predictions for M1 radiative partial widths
(Table~\ref{table:Dudek_LQCD}). First note the reasonably accurate result 
for the ``ground state" ($1S \to 1S$) M1 transition $J/\psi \to \eta_c \gamma$; lattice data and 
experiment are shown together in the rightmost column, and
$\Gamma_{thy.} / \Gamma_{expt.} = 1.36 \pm 0.22$. 
This shows that M1 transitions are reasonably
well reproduced in this LQCD study. The less significant results for 
$2S \to 1S$ $\psi' \to \eta_c \gamma$ involve a well known difficulty; 
since the naive quark model matrix element is zero due to orthogonal
spatial wavefunctions, this is an attempt to extract an intrinsically small amplitude. The most 
interesting result here is the corresponding M1 partial width predicted for the non-exotic 
$J^{PC} = 1^{--}$ charmonium hybrid ``$Y_{hyb.}$", which is 
$\Gamma(Y_{hyb.}\to \eta_c \gamma) = 42 \pm 18 $~keV. By the standards of M1 transitions within $c\bar c$
states this is a large rate, and implies that radiative transitions between hybrids and ordinary 
quarkonium ground states are not significantly suppressed. Even more interesting is the M1 radiative
decay width reported by this reference for the ``classic" hybrid exotic with $J^{PC} = 1^{-+}$, which is
\be
\Gamma(\eta_{c1} (1^{-+}\ {\rm exotic}) \to J/\psi \gamma) \sim 100\ {\rm keV.}
\ee
This is a very large partial width for an M1 transition, 
and is comparable to the known E1 partial widths for transitions
between members of low-lying $c\bar c$ multiplets, such as 
$\psi' \to \chi_{cJ} \gamma$ and  
$\chi_{cJ} \to J/\psi \gamma$. This is exciting indeed, as it implies that the GlueX program 
to search for exotic hybrids using photoproduction at JLAB is well motivated. 
This result may also imply that largely hybrid states (at least with heavier quarks) 
can be identified through their radiative partial 
widths, such as anomalously large M1 widths relative to expectations for conventional quarkonia.
  
In addition to this work on radiative transitions involving hybrids, another very interesting 
new development pursued by the JLAB LQCD group is the determination of the {\it composition} 
of the various states observed on the lattice, through the couplings of these states to a range
of sources (such as radially and orbitally excited $Q\bar Q$ operators, as well as hybrid operators) 
\cite{Dudek:2009qf}. Of course in physical resonances the various basis states 
mix whenever allowed, and the mixing might {\it a priori} be large. However, in their test cases
of higher-mass mesons (with a moderately large quark mass, between $s$ and $c$) the JLAB group finds 
evidence that the basis states are not especially strongly mixed, so that dominantly radial, orbital and 
hybrid excited states can clearly be distinguished, and the pattern of states is broadly consistent 
with naive quark model expectations for the radial and orbital levels. This multisource approach 
to understanding resonances in LQCD is clearly very promising as a technique for improving our 
understanding of the nature of resonances in QCD.

\section{Loops in Charmonium: Unquenching the quark model}

One of the long-standing mysteries in QCD is why the naive valence quark model describes 
the observed hadrons as well as it does. A simple quark potential model, even
a nonrelativistic one, with linear scalar confinement, OGE forces, and a ``constituent quark mass"
describes the spectrum of both mesons (as $q\bar q$ states)
and baryons (as $qqq$ states) rather well, and also gives reasonably good predictions for 
EM, weak and strong transitions between these states. (For $c\bar c$ see for example
Refs.\cite{Eichten:1978tg,Eichten:2005ga,Barnes:2005pb}.) 

The fact that this type of model works reasonably well in describing hadrons
is notoriously to justify. One might have 
expected various corrections to this valence quark model, such as relativistic quark motion, the 
effects of the gluonic degrees of freedom, and mixing between basis states 
(including higher Fock basis states) to completely invalidate this model. 
It is well established nonetheless
that this model gives a reasonable first approximation to the spectrum of hadrons, and in heavy 
quark systems may even now be the most accurate theoretical approach available.
This is one of the principal reasons for the great interest in the light charm-strange mesons
$D_{s0}^*(2317)$ and $D_{s1}(2460)$; they provide an example of a case in which the valence quark potential
model clearly fails to predict masses accurately.

Heavy quark hadrons provide attractive systems for the study of the reasons for the success of the naive 
nonrelativistic valence quark model, since one might expect these corrections to the quark model,
such as relativistic effects and configuration mixing, to be less important. One specific aspect of
configuration mixing that has seen an increase in recent interest is mixing of higher Fock space states,
in particular the effect of an additional light $q\bar q$ pair in heavy meson systems. In one class of
model it is assumed that the important mixing is between the quark model $Q\bar Q$ basis state and 
the continuum of two-meson open flavor color singlet basis states, 
$Q\bar Q \leftrightarrow (Q\bar q)(Q\bar q)$. This involves phenomenological but well-studied 
strong decay amplitudes, such as are usually described using the \tpz model. This type of interaction,
when treated to second order in the decay amplitude, predicts the mass shifts and level of  
$(Q\bar Q) \leftrightarrow (Q\bar Q)$ configuration mixing due to loops of open-flavor mesons, such as
$J/\psi \to D{\bar D} \to J/\psi$ and $\tso (Q\bar Q) \leftrightarrow \tdo (Q\bar Q)$.

Unfortunately, when numerical studies are carried out with physically realistic 
$Q\bar Q \leftrightarrow (Q\bar q)(Q\bar q)$ channel couplings, one finds that the mass shifts 
due to specific individual $(Q\bar q)(Q\bar q)$ channels are so large (and vary considerably
from one $Q\bar Q$ state to another that it is difficult to see how the naive valence quark model, 
which neglects these channel-coupling effects, could ever have been successful.

This conundrum may have been resolved recently in a study 
of coupled-channel effects in charmonium 
reported by Barnes and Swanson \cite{Barnes:2007xu}, who
used the \tpz strong decay model to describe the valence-continuum couplings.
This reference found that although individual intermediate open-flavor continuum channels
made very different contributions to the mass shifts of individual valence states, when a
loop sum was carried out over complete spin multiplets (all $j$ meson states from 
an $n,\ell$ $(Q\bar q) + h.c.$ multiplet), the total mass shift was found to be 
quite similar for every $Q\bar Q$ state within an N,L multiplet. 
This is evident in Table~\ref{table:loop_mass_shifts}, which shows the 
$DD$,
$DD^*$,
$D^*D^*$,
$(c\bar s)$ analog, and summed mass shifts for the 1S,1P and 2S $c\bar c$ states. Clearly the 
individual state mass shifts due to any specific intermediate channel are strongly 
state- and
channel-dependent, but the summed mass shifts are rather similar, and could have been 
largely absorbed in a redefinition of parameters. 

\begin{table}[h]
\begin{tabular}{lr|ccccccc|c} 
\hline
\multicolumn{2}{c|}{Bare $c\bar c$ State}
& 
\multicolumn{7}{c|}{Mass Shifts by Channel, $\Delta \M_i$  (MeV)}
&
\\
Multiplet
&
State\phantom{,,}
&\quad DD &\quad  DD$^*$ &\quad D$^*$D$^*$ 
&\quad  D$_s$D$_s$ &\quad  D$_s$D$_s^*$ &\quad D$_s^*$D$_s^*$ 
& \quad Total \quad & $\;$ P$_{c\bar c}$ \quad \\ 
\hline
1S  
&  $J/\psi(1^3{\rm S}_1) $ 
&  $-23$  & $-83$ & $-132$ 
&  $-21$  & $-76$ & $-123$ 
&  $-457$ 
&  0.69  \\
&  $\eta_c(1^1{\rm S}_0) $ 
&  $\phantom{-}0$  & $-114$ & $-105$ 
&  $\phantom{-}0$  & $-106$ & $-98$ 
&  $-423$ 
&  0.73  \\
\hline
2S 
&  $\psi'(2^3{\rm S}_1) $ 
&  $-27$  & $-84$ & $-126$ 
&  $-19$  & $-70$ & $-113$ 
&  $-440$ 
&  0.51  \\
&  $\eta_c'(2^1{\rm S}_0) $ 
&  $\phantom{-}0$  & $-118$ & $-103$ 
&  $\phantom{-}0$  & $-102$ & $-94$ 
&  $-544$ 
&  0.61  \\
\hline
1P &  $\chi_2(1^3{\rm P}_2) $ 
&  $-40$  & $-105$ & $-144$ 
&  $-33$  & $-88$ & $-111$ 
&  $-521$ 
&  0.49  \\
&  $\chi_1(1^3{\rm P}_1) $ 
&  $\phantom{-}0$  & $-127$ & $-148$ 
&  $\phantom{-}0$  & $-90$ & $-130$ 
&  $-496$ 
&  0.52  \\
&  $\chi_0(1^3{\rm P}_0) $ 
&  $-57$  & $\phantom{-}0$ & $-196$ 
&  $-34$  & $\phantom{-}0$ & $-172$ 
&  $-459$ 
&  0.58  \\
&  $h_c(1^1{\rm P}_1)    $ 
&  $\phantom{-}0$  & $-149$ & $-130$ 
&  $\phantom{-}0$  & $-118$ & $-107$ 
&  $-504$ 
&  0.52  \\
\hline
\hline
\end{tabular}
\caption{Mass shifts and continuum mixing induced in charmonium states by open-charm 
$(c\bar q)(q\bar c)$ meson loops, as found by Ref.\cite{Barnes:2007xu}. Note that the 
mass shifts due to individual vary widely, but the sum is roughly state independent.
Also note (final column) that the resulting physical states have rather large continuum
components. ($P_{c\bar c} = |\langle\Psi|c\bar c\rangle|^2$ is the overlap of the physical
state with the original $c\bar c$ valence state.)} 
\label{table:loop_mass_shifts} 
\end{table}

Even more remarkably, given certain rather mild constraints on the members of these multiplets 
(identical spatial wavefunctions within each N,L $Q\bar Q$ multiplet, and separately within
each $n,\ell$ $Q\bar q$ multiplet, 
initially degenerate valence states within each multiplet,
and \tpz decay couplings), Ref.\cite{Barnes:2007xu} found that this was an {\it exact}
result: An initially degenerate valence multiplet remains degenerate when coupled-channel
(hadron loop) effects are included. (The mass shift is the same for all $Q\bar Q$ 
states in a multiplet.) This explains how loop effects on the hadron spectrum could be 
both large {\it and} hidden; they were subsumed in an overall mass shift 
that is approximately state independent.

One can also prove two other related results, regarding loop-induced mixing of different 
valence basis states through loops (which cancels in this limit) and strong widths (which are 
all identical in this limit, which imposes identical phase space for decays). 
This leads us to the ``{\bf Three Laws of Loopotics}" (with apologies to Asimov~\cite{Asimov}), 
which are that {\it given the constraints specified above}, 
the following results hold for the $Q\bar Q$ states $\{ \A\}$ in a given N,L multiplet: 
\vskip 0.5cm

\begin{enumerate}
\item
{\bf The mass shifts due to hadron loops for all states $\{ \A\}$ in a given N,L multiplet are equal.}
\item
{\bf Their strong (open-flavor) total widths are also equal.}
\item
{\bf The configuration mixing amplitude $a_{fi}$ between
any two valence basis states $i$ and $f$ due to hadron loops vanishes if
$L_i \neq L_f$ or $S_i \neq S_f$.}
\end{enumerate}
\vskip 0.5cm

These results have since been extended to more general decay models by Close and Thomas
\cite{Close:2009ii}. We also note that a similar result regarding common loop mass shifts was 
previously reported by Tornqvist \cite{Tornqvist:1979hx}. Recent studies of loop
effects on charmed hadrons or charmonia include work by
Baru {\it et al.} \cite{Baru:2010ww},
Gamermann {\it et al.} \cite{Gamermann:2009uq},
Guo, Krewald and Meissner \cite{Guo:2007up},
Huang and Kim \cite{Hwang:2004cd}, 
Simonov and Tjon \cite{Simonov:2004ar}, and
van Beveren and Rupp \cite{vanBeveren:2006st}.
Since the continuum components of physical hadrons are typically found to be relatively large,
it would be interesting to identify a ``smoking gun" experimental measurement that is sensitive
to loop components, such as a radiative transition in which the photon couples much more strongly 
to a $Q\bar q$ loop meson than to the heavy-quark $Q\bar Q$ valence state.  

\eject

\section{PANDA Cross Sections and BES}

The PANDA project at GSI \cite{PANDA} proposes to produce charmonia, 
and ultimately $J^{PC}$-exotic charmonium hybrids, using $p\bar p$ annihilation.
Although $p\bar p$ annihilation has been used previously to study charmonium,
by E760 and the follow-on experiment E835 at Fermilab, these earlier studies used direct 
s-channel annihilation ($p\bar p \to c\bar c \to $~hadrons), which only produces non-exotic 
$J^{PC}$ states. Exotics will instead require associated production, 
in which a $J^{PC}$-exotic charmonium hybrid $H_{c\bar c}$ (for example)
recoils against another hadronic system, such as a meson $m$; $p\bar p \to m + H_{c\bar c}$. 

The PANDA project will only be feasible if these cross sections are sufficiently large. It is therefore 
somewhat unsettling that very little is known about these cross sections experimentally; the entire (published) 
data set on these associated charmonium cross sections consists of a few cross section measurements 
for the process $p\bar p \to \pi^0 J/\psi$ near the $h_c$ mass, which were taken by E760 and E835 
for background estimates. 
(Data from E760 and E835 are discussed in Ref.\cite{Andreotti:2005vu}, which however does not quote
acceptance-corrected cross sections for this reaction.) These earlier Fermilab measurements 
do provide a cross section scale ($\sim 100$~pb), although we have no indication from this data of how 
these associated charmonium cross sections might vary with the choice of the light meson $m$, 
the charmonium state, or the $\sqrt{s}$ of the reaction.

There have been several calculations of these cross sections using results from related processes, 
notably charmonium decays of the type $(c \bar c) \to p\bar p m$, which have now been observed in 
several channels, and are related to the PANDA reactions by crossing. 
The earliest of these calculations (Gaillard {\it et al.} \cite{Gaillard:1982zm}, from the LEAR era) 
introduced a hadronic pole model, and discussed angular distributions 
in $p\bar p \to \pi^0 J/\psi$ at one energy.
Lundborg {\it et al.} \cite{Lundborg:2005am} next derived a crossing relation, and used
$\Gamma(J/\psi \to p\bar p \pi^0)$ to predict $\sigma(p\bar p \to \pi^0 J/\psi)$ in a simple 
constant-amplitude model; the results were about a factor of 2-3 larger than the E760 cross section. 
The Gaillard {\it et al.} hadronic pole model for $p\bar p \to \pi^0 J/\psi$ 
was next generalized to other channels by Barnes and Li \cite{Barnes:2006ck}, 
who reported cross section results for a wide range of 
PANDA ($p\bar p \to m + (c\bar c)$) reactions. Other interesting related results include 
evidence for a Pauli term in the $J/\psi p\bar p$ hadronic vertex \cite{Barnes:2007ub} 
(suggested by BES data for the angular distribution of $e^+e^- \to J/\psi \to p\bar p$) and 
(note added in proof) the possibility of extracting $NNm$ meson-nucleon strong couplings 
directly from the decays $(c\bar c)  \to p \bar p m$ at BES \cite{Barnes:2010yb}.
 
\begin{figure}[ht]
\includegraphics[width=2.5in]{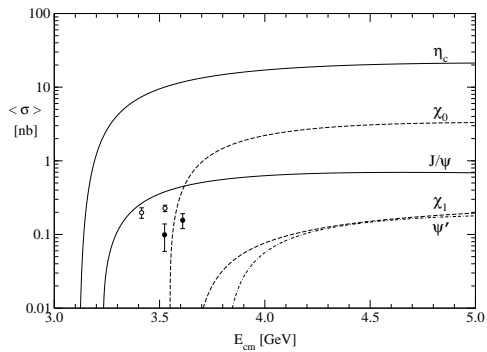}
\caption{\label{fig:csecs} Theoretical ($p\bar p \to m + (c\bar c)$)
and experimental ($p\bar p \to \pi^0 J/\psi$, from E760 and E835) 
PANDA type cross sections, from Ref.\cite{Barnes:2006ck}.}
\end{figure}

A comparison between theory (Ref.\cite{Barnes:2007ub}, for a range of $p\bar p \to m + (c\bar c)$
cross sections), and experiment (E760 and E835, for $p\bar p \to \pi^0 J/\psi$ only) is shown in
Fig.\ref{fig:csecs}. Note that the cross sections for producing the $\eta_c$ and $\chi_0$ are
much larger than the other cases considered, due to their large $(c\bar c)p\bar p$ couplings, which
are inferred from the $(c\bar c)\to p\bar p$ partial widths. This may imply that $c\bar c$ states 
couple most strongly to $p\bar p$ through $gg$ intermediate states, 
so that PANDA would find the largest associated production cross sections 
for charmonia or charmonium hybrids having $gg$ quantum numbers. 

\vskip -1cm
\section{Deep Charm Molecules}

Another topic that has been affected by the renaissance in charm spectroscopy is the subject of hadronic 
molecules. The discovery of the X(3872) in $J/\psi\pi^+\pi^-$ in B decays by Belle \cite{Choi:2003ue}
has given us our strongest candidate for a meson-meson molecular state, since its mass and quantum numbers 
are consistent with a weakly bound $D^0D^{*0}$ ($h.c.$ implicit) pair, but are inconsistent 
with expectations for a conventional $c\bar c$ state. Theoretical models suggest that this 
$J^{PC} = 1^{++}$ $D^0D^{*0}$ 
system experiences an attractive force due to one pion exchange that is just strong enough to form 
a weakly bound state (perhaps also requiring additional, subdominant interactions).

Although the system is an S-wave hadron pair, which is clearly favored for binding, Close and Downum 
\cite{Close:2009ag} recently
made the interesting observation that the two hadronic vertices involved in this one pion exchange require
a P-wave $D\pi$ pair. (e.g. $D{\bar D}^*(1^-) \to D[{\bar D}(0^-)\pi (0^-)]_P \to [D(0^-)\pi (0^-)]_P{\bar D} 
\to D^*(1^-){\bar D}$.) They note that processes in which the pion is emitted and absorbed in an 
S-wave meson-pion system should experience much
stronger forces in general, and (when attractive) may form more deeply bound charm meson molecules. This S-wave
requirement suggests consideration of molecules of meson pairs with the same J and opposite P, such as 
$D^*D_1$. This idea has been extended to other heavy flavor systems by Close, Downum and Thomas 
\cite{Close:2010wq}.

The binding energies calculated for the S-wave pion systems are strongly dependent on the
short-distance behavior assumed for pion exchange, so precise values for the predicted binding energies
cannot yet be quoted; Close and Downum estimate that these ``deep charm molecules" might be {\it ca.} 
100~MeV below threshold, hence near 4.2~GeV for $D^*D_1$, which suggests that the Y(4260) be considered 
a candidate. A search for $D{\bar D} \pi\pi\pi$ decay modes of the Y(4260) is motivated by this possibility, 
since a $D^*D_1$ molecule would likely decay significantly through constituent decay, $D^*\to D\pi$ and 
$D_1\to D\pi\pi$. (Again an antiparticle $h.c.$ label is implicit.)

\vskip -1cm
\section{Charmiscelleny}

We next consider two short items in charmonium spectroscopy which do not fit into
the major categories discussed above, relating to the $\psi(4415)$ and the $Z(3930)$.

The $\psi(4415)$ was for decades the highest mass charmonium state known, and in $c\bar c$ potential models
is most often given the assignment $4\, {}^3$S$_1$. Although this state appears as a clear peak in $R$, until
very recently no exclusive decay modes had been identified, so the PDG (also for decades) listed this state 
as decaying dominantly to ``hadrons", which one might have considered a safe assumption. 
This situation has finally improved with the observation of an exclusive $\psi(4415)$ 
decay mode by Belle~\cite{Pakhlova:2007fq}; it is seen in 
$D_2^*(2460)D$, with a combined branching fraction for
$B(\ps\to\ddt)\times B(\dt\to\dpi)$ of $\approx 10-20\%$ (depending on the $\psi(4415)$ parameters assumed).
On considering other charge states and additional channels such as $D_2^*(2460)\to D^*\pi$,
this appears roughly consistent with the theoretical prediction
of Ref.\cite{Barnes:2005pb}, which anticipated that this was the second largest mode of a $4\, {}^3$S$_1$ 
$c\bar c$ $\psi(4415)$, with a branching fraction of $\approx 30\%$. As the largest mode is predicted to be
$DD_1(narrow)$ (in a relative D-wave!), with a $\approx 40\%$ branching fraction, a search for this mode 
is clearly of interest.
\vskip 0.5cm

And finally; the $Z(3930)$ seen by BABAR~\cite{Uehara:2005qd} in two-photon fusion, 
$\gamma\gamma \to Z(3930) \to DD$, is now widely accepted as a $2\, {}^3$P$_2$ $c\bar c$ state 
``$\chi_{c2}(3930)$", due largely to the agreement
with the expected angular distribution for J=2, and to the plausible mass for a 2P $c\bar c$ state. It would be 
edifying to complete this picture through the identification of this state in 
$\gamma\gamma \to \chi_{c2}(3930) \to DD^*$;
$DD^*$ is the only other open-charm mode available to this state, and the predicted relative branching fraction
to $DD^*$~\cite{Barnes:2005pb} is a reasonably large $B(\chi_{c2}(3930) \to DD^* / DD) \approx 0.35$.

\vskip -1cm
\section{Conclusions}

In this invited talk I have reviewed some of the more recent developments in the theory of charmonium and 
related states; specific topics discussed included double charmonium production, LQCD and charmonium, 
open-charm hadron loops, charmonium production cross sections at PANDA, deeply bound charm molecules, 
and two additional (recent and possible future) measurements relating to $c\bar c$ candidates. 
It is evident that the spectroscopy of charmonium and 
related states is a remarkably active and interesting field of research in hadron physics. 
With the development of the BES-III and PANDA facilities, this happy situation should continue 
for many years to come. 

\vskip -1cm
\section{Acknowledgments}
I am grateful to Paul Eugenio and the conference organizers for their kind invitation to present this talk 
at HADRON09. Research sponsored in part by the Office of Nuclear Physics, U.S. Department of Energy.

\vfill\eject

\end{document}